\documentclass[useAMS,usenatbib]{mn2e}
\usepackage{graphicx}
\usepackage{txfonts}
\usepackage{natbib}

\def\beq#1{\begin{equation}\label{#1}}
\def\eeq{\end{equation}}
\def\beqa#1{\begin{eqnarray}\label{#1}}
\def\eeqa{\end{eqnarray}}

\def\myfrac#1#2{\left(\frac{#1}{#2}\right)}

\def\comment#1{\relax}

\title[Bright Flares in SFXTs]{Bright Flares in Supergiant Fast X-ray Transients}
\author[N. Shakura et al.] {N. Shakura$^1$
\thanks{E-mail: nikolai.shakura@gmail.com, kpostnov@gmail.com},
K. Postnov$^1$,  L. Sidoli$^2$, A. Paizis$^2$\\
$^{1}$
Sternberg Astronomical Institute, Moscow M.V. Lomonosov State University, Universitetskij pr., 13, 119992, Moscow, Russia
\\	
$^{2}$INAF, Istituto di Astrofisica Spaziale e Fisica Cosmica, Via E.\ Bassini 15,   I-20133 Milano,  Italy}

\begin{document}

\date{Received ... Accepted ...}
\pagerange{\pageref{firstpage}--\pageref{lastpage}} \pubyear{2012}

\maketitle

\label{firstpage}

\begin{abstract}
At steady low-luminosity states, Supergiant Fast X-ray Transients (SFXTs) can be 
at the stage of quasi-spherical settling accretion onto slowly rotating 
magnetized neutron stars from the OB-companion winds.
At this stage, a hot quasi-static shell is formed above the magnetosphere, 
the plasma entry rate into magnetosphere is controlled by (inefficient) 
radiative plasma cooling, and the accretion rate onto the neutron star is suppressed 
by a factor of $\sim 30$ relative to the Bondi-Hoyle-Littleton value. 
Changes in the local wind velocity and density due to, e.g., clumps, can only slightly increase the mass accretion rate
(a factor of $\sim 10$) bringing the system into the Compton cooling dominated regime and led to the production of moderately bright flares ($L_x\lesssim 10^{36}$~erg/s).
To interpret the brightest flares ($L_x>10^{36}$~erg/s) displayed by the SFXTs within the quasi-spherical settling accretion regimes,
we propose that a larger increase in the mass accretion rate can be produced by 
sporadic capture of magnetized stellar wind plasma. At sufficently low accretion rates, 
magnetic reconnection 
can enhance the magnetospheric plasma entry rate, resulting in copious production of X-ray photons,
strong Compton cooling and ultimately in unstable 
accretion of the entire shell. 
A bright flare develops on the free-fall time
scale in the shell, and the typical energy released in an SFXT bright flare corresponds to 
the mass of the shell. This view is consistent with 
the energy released in SFXT bright flares ($\sim 10^{38}-10^{40}$~ergs),
their typical dynamic range ($\sim 100$), and with the observed dependence of these characteristics on the average unflaring X-ray luminosity of SFXTs. 
Thus the flaring behaviour of SFXTs, as opposed to steady HMXBs, may be primarily related to 
their low X-ray luminosity allowing sporadic magnetic reconnection to occur during 
magnetized plasma entry into the magnetosphere.

\end{abstract}

\begin{keywords}
accretion - pulsars:general - X-rays:binaries
\end{keywords}

\section{Introduction}
\label{intro}

%observations:
%\cite{2008A&A...484..783C}
%\cite{2008A&A...484..801R}
%\cite{2014A&A...562A...2R}

%reviews:
%\cite{2011AdSpR..48...88S}

%likely models:
%\cite{2007AstL...33..149G}
%\cite{2009MNRAS.398.2152D}
%\cite{2008ApJ...683.1031B} 

Supergiant Fast X-ray Transients (SFXTs) are a subclass of 
%heavily obscured 
High-Mass X-ray Binaries (HMXBs) 
associated with early-type supergiant companions (\citealt{Pellizza2006}, \citealt{Chaty2008}, \citealt{Rahoui2008}),
and characterized by sporadic, short and bright X--ray flares 
reaching peak luminosities of 10$^{36}$--10$^{37}$~erg~s$^{-1}$.
Most of them were discovered by INTEGRAL (\citealt{2003ATel..176....1M, 2003ATel..190....1S, 
2003ATel..192....1G}, \citealt{Sguera2005}, \citealt{Negueruela2005}).
They show high dynamic ranges (between 100 and 10,000, depending on the specific source; 
e.g. \citealt{Romano2011, 2014A&A...562A...2R}) and their X-ray spectra in outburst are very similar to accreting pulsars in HMXBs.  
In fact, a half of them have measured neutron star (NS) spin periods similar to those observed 
from persistent HMXBs (see \citealt{2012int..workE..11S} for a recent review).

The physical mechanism driving their transient behavior, related with the accretion by the compact
object of matter from the supergiant wind, has been discussed by several authors
%(Grebenev & Sunyaev 2007; Ducci et al. 2009; Bozzo et al. 2008), 
and it is still a matter of debate, as some of them require particular properties of the compact objects hosted in these systems 
(\citealt{2007AstL...33..149G}, 
%\citealt{2009MNRAS.398.2152D},
\citealt{2008ApJ...683.1031B}), 
and others assume
peculiar clumpy properties of the supergiant winds and/or orbital characteristics 
(\citealt{zand2005,Walter2007,
Sidoli2007,
Negueruela2008,2009MNRAS.398.2152D,Oskinova2012}).

Recently, a new approach has been proposed from the observational point of view:
cumulative luminosity distributions of SFXTs emission have been
built for the first time in hard X--rays using the long based {\it INTEGRAL}
public archive \citep[$\sim$9 years of IBIS data,][]{PaizisSidoli2014}. This work enables a self consistent overview of all
currently known SFXTs, with quantitative information on source duty cycles,
range of variability and shape of the hard X-ray luminosity distributions,
especially compared to persistent systems.

In this paper we propose an alternative mechanism for bright flares in SFXTs, which
is based on an instability of the quasi-spherical shell around the magnetosphere 
of a slowly rotating NS. Such shell must be formed at low mass accretion rates
$\lesssim 4\times 10^{16}$~g s$^{-1}$ (low accretion X-ray luminosities 
$\lesssim 4\times 10^{36}$~erg s$^{-1}$)  in quasi-spherical accretion from stellar wind of the optical companion \citep{2012MNRAS.420..216S}. In this model, the actual accretion rate 
through the magnetosphere is controled by the plasma cooling.
(Compton at higher luminosities or radiative at lower ones). 
%and can be much smaller than the Bondi-Hoyle-Littleton 
%accretion rate determined by the stellar wind properties and NS
%orbital velocity. 
The switch from Compton to inefficient radiation cooling 
is expected to occur at luminosities $\lesssim 10^{35}$~erg~s$^{-1}$ 
\citep{2013MNRAS.428..670S}, resulting in actual accretion rates typically 
a factor of $\sim 30$ smaller than the maximum possible Bondi-Hoyle-Littleton 
rate determined by the stellar wind properties and NS
orbital velocity. Variations in the stellar wind velocity and density 
are translated into plasma density changes near the magnetosphere
and hence, via cooling, into the X-ray luminosity variations. 
Above some accretion rate, the restructure of
X-ray beam pattern from NS can bring the source into more effective 
Compton cooling regime, resulting in a moderate flare with $L_x\lesssim 10^{36}$~erg~s$^{-1}$. 
To explain the bright SFXT flares, however, a much more efficient 
plasma entry rate is required, signaling an instability in the shell.  
We suggest that the instability can be triggered by the sudden increase in the mass
accretion rate through the magnetosphere due to reconnection of the  
large-scale
magnetic field sporadically carried by the stellar wind of the
optical OB companion.
%brought into the shell from the magnetized stellar wind of the optical OB supergiant.     
We show that this instability can reproduce the observed energy released in bright
flares of SFXTs, as well as the dynamic range of the flares and their dependence on the 
mean unflaring X-ray luminosity in different sources. 
The bright flares due to the proposed mechanism can occur on top of smooth 
variation of mass accretion rate due to, for example, orbital motion of the neutron star in a binary 
system.   
%%%%%%%%%%%%%%%%%%%%%%%%%%%%%%%%%%%%%%%%%%%%%%%%%%%%%%%%%%%%%%%%%%%%%%%% Fig 
\begin{figure}
\begin{center}
\centerline{\includegraphics[width=0.5\textwidth]{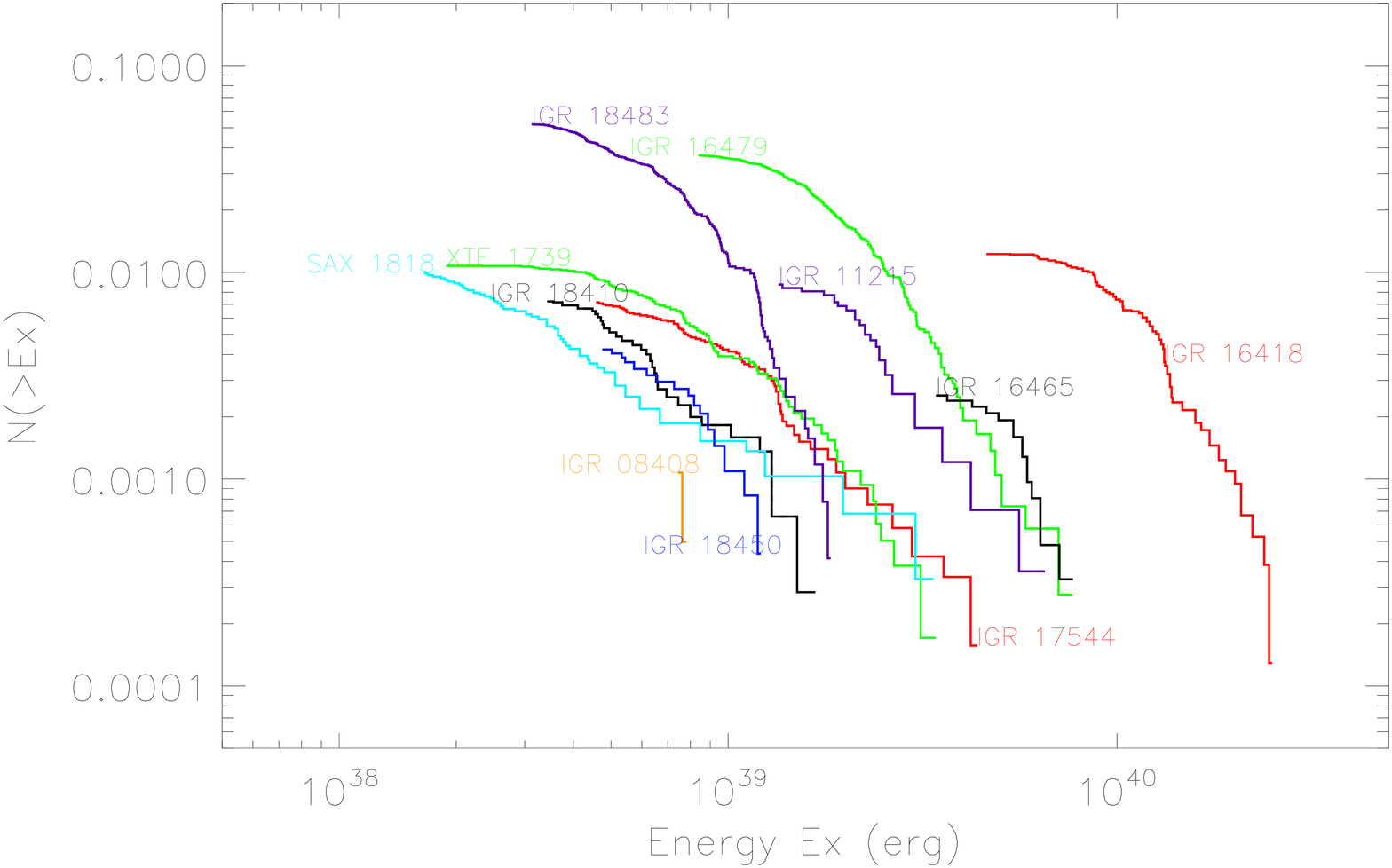}} 
\caption{Cumulative distribution of the total energy released in SFXT flares ({\it INTEGRAL}/IBIS data, 
17--50\,keV). The fraction of time (\emph{y-axis}) spent by SFXTs above a certain energy value (\emph{x-axis}) is given.}
\label{fig:energy}
\end{center}
\end{figure}
%%%%%%%%%%%%%%%%%%%%%%%%%%%%%%%%%%%%%%%%%%%%%%%%%%%%%%%%%%%%%%%%%%%%%%%%

      %%%%%%%%%%%%%%%%%%%%%%%%%%%%%%%%%%%%%%%%%%%%%%%%%%%%%%%%%
        \section{Energy released in SFXT flares}\label{Fig1}
        %%%%%%%%%%%%%%%%%%%%%%%%%%%%%%%%%%%%%%%%%%%%%%%%%%%%%%%%%
%
%Updated version of this section (please completely remove the previous one)
%
%
Given the rare, sporadic and intermittent character
of the bright flares in SFXTs (usually, single outbursts are separated by several months of either much fainter X--ray emission or quiescence), 
the exploitation of very long based archival data is the best strategy to obtain
reliable quantitative information about the source duty cycles, X--ray range of variability, shape of the luminosity distributions.

The temporal profiles of SFXT flares are usually very complex and it is not straightforward to define or disentangle an X--ray flare.
However, since time scales of flare durations are roughly consistent with  {\it INTEGRAL}  pointing 
durations (order of ks), we consider the single pointing detections as a good representation of SFXT flares. 

The energy released during SFXT flares, as observed by {\it INTEGRAL}, is shown in Figure~\ref{fig:energy}. The plot is an overview of all known SFXTs\footnote{IGR\,J17544--2619, IGR\,J16418--4532, IGR\,J16479--4514, IGR\,J16465--4507, SAX\,J1818.6--1703,  IGR\,J18483--0311, XTE\,J1739--302, IGR\,J08408--4503, IGR\,J18450--0435, IGR\,J18410--0535, IGR\,J11215--5952}. The data selection and analysis is discussed in detail in \cite{PaizisSidoli2014}, together with the assumed distances and relevant references, so we refer the reader to that paper for the technical details.
In Figure~\ref{fig:energy}, for each source we built the complementary cumulative distribution function of the obtained X--ray flare energies. In each point of these functions, at a given energy E$_{X}$ (erg), the sum of all events (detections) with an energy larger than E$_{X}$ are plotted. The energy released in each flare is obtained by multiplying the flare luminosities as obtained in \cite{PaizisSidoli2014}, Figure~3, by the duration of the pointings.

Each curve has been normalised to the total exposure of the source fields observed by IBIS, hence they can be directly compared: the energy released during flares by SAX\,J1818.6--1703 is $>1.7\times10^{38}$~ergs for about 1\% of the total observing time (7.3\,Ms), whereas the energy released in IGR\,J18483--0311 flares is higher than $3.1\times10^{38}$ ergs for 5\% of the total observing time (6.0\,Ms). The lower end of the curves (i.e. towards dimmer flares) is limited by the IBIS/ISGRI sensitivity in a single pointing.

%\begin{figure*}
%\includegraphics[width=\textwidth]{LogNlogE_17-50keV_5s12d_all_EN_SFXT.eps}
%\caption{Distribution of total energy released in SFXT outbursts}
%\label{f:sfxt_en}
%\end{figure*} 

\section{Model of SFXT bright flares}
\label{sec:model}

\subsection{Settling subsonic accretion onto slowly rotating NS}

There can be two distinct cases of quasi-spherical accretion onto slowly rotating magnetized NS. 
The classical 
Bondi-Hoyle-Littleton 
accretion takes place when the shocked matter rapidly cools down (via 
Compton cooling), and the matter freely falls toward the NS magnetosphere.  
%(see Fig. \ref{f:1}). 
A shock in the accreting plasma is formed at some distance above the magnetosphere. 
Close to the magnetopause, the shocked matter rapidly cools down 
and enters the magnetopshere via Rayleigh-Taylor instability \cite{1976ApJ...207..914A}.
The magnetospheric
boundary is characterized by the Alfv\'en radius $R_A$, which can be
calculated from the balance of the ram pressure of the 
infalling matter and the magnetic
field pressure at the boundary. 
This regime of quasi-spherical accretion is realized in 
bright X-ray pulsars with $L_x>4\times 10^{36}$~erg s$^{-1}$ \citep{2012MNRAS.420..216S}. 

If the shocked matter remains hot (when plasma cooling time 
is much longer than the free-fall time, $t_{cool}\gg t_{ff}$), 
a hot quasi-static shell with almost iso-angular momentum distribution 
forms above the magnetosphere. The subsonic 
(settling) accretion sets in. 
The shell mediates the
angular momentum transfer from/to the NS magnetosphere via viscous stresses
due to convection. In this regime, the mean radial velocity 
of matter in the shell $u_r$ is smaller than the free-fall velocity $u_{ff}$: 
$u_r=f(u)u_{ff}$, $f(u)<1$, and is determined by the plasma cooling rate 
near the magnetosphere: 
$f(u)\sim [t_{ff}(R_A)/t_{cool}(R_A)]^{1/3}$. Here the actual mass accretion rate
onto NS can be significantly smaller than the Bondi mass accretion rate, $\dot M_a=f(u) \dot M_B$.  
The settling accretion is expected to occur 
at $L_x<4\times 10^{36}$~erg s$^{-1}$ (see \citealt{2012MNRAS.420..216S, 2013arXiv1302.0500S, 2014EPJWC..6402001S} for the detailed derivation, discussion and applications to
spin-up/spin-down of slowly rotating low-luminosity X-ray pulsars).

Different regimes of plasma cooling near the magnetosphere (radiative or Compton)
can take place. 
The switch-off of the Compton cooling 
is expected with decreasing X-ray luminosity, which can be related to  
X-ray beam generated near the NS surface switching 
from fan-like to pencil-like form, as discussed in \cite{2013MNRAS.428..670S} with regard to 
temporal appearance of low-luminosity 
'off' states in Vela X-1. The pulse profile phase change associated with X-ray beam switching below some critical luminosity, as observed in Vela X-1, seems to be suggested by an $XMM-Newton$ observation of the SFXT IGR~J11215--5952 (see Fig.~3 in \cite{Sidoli2007}),
corroborating  
the subsonic accretion regime with radiative plasma cooling at low X-ray luminosities in SFXTs as well. 
 
\subsection{Energy released in bright flares}

Fig. \ref{fig:energy} suggests that the typical energy released in a SFXT bright flare is about 
$10^{38}-10^{40}$~ergs, varying by one order of magnitude for different
sources. That is, the mass fallen onto the NS
in the typical bright flare varies from $10^{18}$~g to around $10^{20}$~g. 
(assuming an X-ray emission efficiency of about 0.1). 

The typical X-ray luminosity outside outbursts in SFXTs is about 
$L_{x,low}\simeq 10^{34}$~erg s$^{-1}$ \citep{2008ApJ...687.1230S},
and below we shall normalise the luminosity to this value, $L_{34}$. 
At these small X-ray luminosities, plasma entry rate into magnetosphere is controlled 
by radiative plasma cooling.
Next, it is convenient to normalise the typical stellar wind velocity from hot OB-supergiants $v_w$ 
to 1000~km~s$^{-1}$ (for orbital periods about few days and larger the NS orbital velocities can be neglected compared to the stellar wind velocity from the OB-star), so that the Bondi gravitational capture radius is $R_B=2GM/v_w^2=4 \times 10^{10}v_{8}^{-2}$~cm 
for a fiducial NS mass of $M_x=1.5 M_\odot$.

Consider a quasi-static shell hanging over the magnetosphere around the NS, with the magnetospheric accretion rate being controlled by the radiative plasma cooling.
We denote the actual steady-state accretion rate as $\dot M_a$ so that the observed X-ray steady-state luminosity  is $L_x=0.1\dot M_a c^2$. Then from the theory of subsonic 
quasi-spherical accretion \citep{2012MNRAS.420..216S} we know that the factor $f(u)$ (the ratio of the actual velocity of plasma entering the magnetosphere due to the Rayleigh-Taylor instability 
to the free-fall velocity at the magnetosphere,
$u_{ff}(R_{A})=\sqrt{2GM/R_A}$) reads \citep{2013MNRAS.428..670S,2014EPJWC..6402001S}
%\beq{fucomp}
%f(u)_{Comp} \simeq 0.041 L_{34}^{4/11}\mu_{30}^{-1/11}
%\eeq
\beq{furad}
f(u)_{rad} \simeq 0.036 L_{34}^{2/9}\mu_{30}^{2/27}\,.
\eeq
Here $\mu_{30}=\mu/10^{30}$~G~cm$^3$ is the NS dipole magnetic moment;
note a very weak dependence on the NS magnetic field.

The shell is quasi-static (likely convective), unless something triggers
much faster matter fall through the magnetosphere  (see below the possible reason). 
It is straightforward to calculate the mass of the shell 
using the density distribution $\rho(R)\propto R^{-3/2}$ 
\citep{2012MNRAS.420..216S}. Using the mass continuity equation to eliminate the density above the magnetosphere, we readily find 
\beq{deltaM}
\Delta M \approx \frac{2}{3} \frac{\dot M_a}{f(u)}t_{ff}(R_B)\,.
%\a\dot M_B t_{ff}(R_B)\,.
\eeq
Note that this mass can be expressed through measurable values
$L_{x,low}$, $\mu_{30}$ and the (directly unobserved) stellar wind
velocity at the Bondi radius $v_w(R_B)$. Using Eq. (\ref{furad}) for  
radiative plasma cooling, we obtain
%\beq{deltaMcomp}
%\Delta M_{Comp}% = \frac{2}{3} \frac{\dot M_a}{f(u)} \frac{R_B^{3/2}}{\sqrt{2GM_x}}=
%\frac{2}{3} \frac{\dot M_a}{f(u)}  \frac{2GM_x}{v_w^3} 
%\approx 7\times 10^{17} [g] L_{34}^{7/11} v_8^{-3}\mu_{30}^{1/11}\,.
%\eeq
\beq{deltaMrad}
\Delta M_{rad} 
%= \frac{2}{3} \frac{\dot M_a}{f(u)} \frac{R_B^{3/2}}{\sqrt{2GM_x}}=
%\frac{2}{3} \frac{\dot M_a}{f(u)}  \frac{2GM_x}{v_w^3} 
\approx 8\times 10^{17} [g] L_{34}^{7/9} v_8^{-3}\mu_{30}^{-2/27}\,.
\eeq
The simple estimate (\ref{deltaMrad}) shows that for the likely wind velocity 
near the NS about 500~km~s$^{-1}$ the \emph{typical} mass of the hot 
magnetospheric shell is around $10^{19}$~g, 
corresponding to $10^{39}$~ergs released in a flare in which all the matter from the shell
is accreted onto the NS, as observed (see Fig. \ref{fig:energy}). 
Clearly, the wind velocity variations in different sources by factor $\sim 2$  
would produce a one-order-of-magnitude spread in $\Delta M$ observed in bright SFXT flares.

%%%%%%%%%%%%%%%%%%%%%%%%%%%%%%%%%%%%%%%%%%%%%%%%%%%%%%%%%%%%%%%%%%%%% Fig 2
\begin{figure}
%\begin{center}
\includegraphics[width=0.45\textwidth]{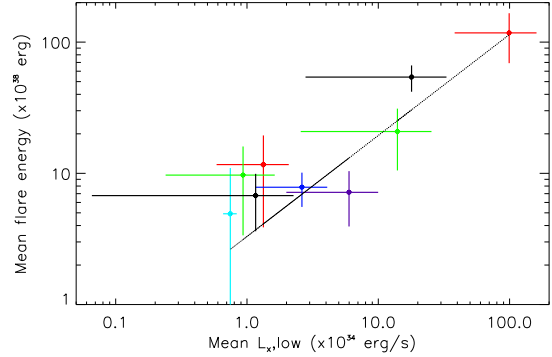}
\caption{The mean energy released in bright flares (17 -- 50 keV, data
from \citet{PaizisSidoli2014}) versus average
\textit{INTEGRAL}/IBIS source luminosity.
The average luminosities have been obtained by the source fluxes measured in time-averaged maps 
\citet{krivonos2012} (17--60\,keV, IBIS survey over 9 years, 
nine SFXTs). The same distances used to derive our flare 
energies have been considered for the estimates of the average 
luminosities. Sources are colored as in Fig.~1. 
Errors shown are explained in text. The \textit{x-axis}
is in units of $10^{34}$~erg~s$^{-1}$,
the \textit{y-axis} is in units of $10^{38}$~ergs.
The straight line gives the formal rms linear fit with the slope 
$0.77 \pm 0.13$. }
%
%
%The mean energy of SFXT bright flares $0.1\Delta M c^2$ as a function of low (unflaring) X-ray luminosity for sources from Fig. \ref{fig:energy} with errors shown as explained in text. The straight line gives the formal rms linear fit with the slope 
% $0.77 \pm 0.13$.} \label{fig:dM}
%}
%
\label{fig:dM}
%}
%\end{center}
\end{figure}

In Fig. \ref{fig:dM} we show the mean energy of SFXT bright flares $\Delta E=0.1\Delta M c^2$ as a function of low (unflaring) X-ray luminosity for sources from Fig. \ref{fig:energy}.
The low (unflaring) X-ray luminosity (\emph{x-axis}) has been taken from Krivonos et al. (2012),
where a nine year time-averaged source flux in the 17--60\,keV band is given for each source 
(with the exception of two, IGR~J11215--5952 and IGR~J08408--4503)
\footnote{We chose these data as a representation of the non-flaring activity to be compared to our
flare detections (Figure~1) because our energy range (17--50\,keV) and data sample ($\sim$9 years) 
basically overlap (most of the emission from SFXTs is below 50\,keV), while \citet{Bird2010}
adopt a slightly different energy range (20--40\,keV band) and include less data}.
The uncertainties on the low luminosities  
include both the statistical errors on source fluxes, as reported in  Krivonos et al. (2012),
and the known SFXTs distances and their uncertainties as reported by  Paizis \& Sidoli (2014). 
For clarity, we assumed an uncertainty of 1~kpc for all SFXTs except for SAX~J1818.6-1703 (for which we used 0.1~kpc), 
IGR~J16465-4507 and IGR~J16418-4532 (4~kpc),   IGR~J16479-4514 (2~kpc) and IGR~J18410-0535 (1.5 kpc).
The errors on the \emph{y-axis} in Fig.~2 are based on the standard deviation of the bright flare energies,  
giving a conservative idea of the range of variability of bright flare energy as observed by $INTEGRAL$.
The formal rms fit to these points gives the dependence of 
$\Delta E_{38}=(3.3\pm 1.0)L_{34}^{0.77\pm  0.13}$. 
This exactly corresponds to radiative cooling regime 
(see Eq. (\ref{deltaMrad}), as expected.  
%though the Compton cooling dependence cannot be excluded within the present errors. 
Comparison with coefficient in formula  (\ref{deltaMrad}) suggests $v_8\sim 0.62$,  
the typical wind velocities observed in HMXBs.

What can trigger the SFXT flaring activity? As noted in 
\cite{2013MNRAS.428..670S},
if there is an instability leading to rapid matter fall through the magnetosphere, 
a lot of X-ray photons produced near the NS surface should 
rapidly cool down the plasma, further increasing the plasma fall velocity
through magnetosphere and the ensuing accretion NS luminosity. Therefore, in a bright flare 
the entire shell can fall onto the NS on the free-fall time scale from the outer 
radius of the shell $t_{ff}(R_B)\sim 1000$~s. Clearly, the shell will be replenished by
new wind capture, so the flares will repeat as long as the 
rapid mass entry rate into the magnetosphere is sustained.

\subsection{Magnetized stellar wind as the flare trigger}
\label{sec:magwind}

Observations suggest that about $\sim 10\%$ of hot OB-stars have magnetic fields
up to a few kG (see \cite{2009ARA&A..47..333D, 2013arXiv1312.4755B} for a review and discussions).
The fields of massive stars affect their winds
rather than their near-surface distribution of chemical elements
and are capable of reshaping the wind by forcing
the escaping plasma to follow the field lines. This results in a complex variable
wind structure around a rotating magnetized star (see, for example, 
the analysis of X-ray observations and MHD-simulations of the wind
around young magnetic O-star $\theta^{1}$ Orionis C in \cite{2005ApJ...628..986G}).
According to such calculations, within the Alfv\'en radius of the star,
where the wind flow is dominated by the large-scale magnetic field of the star, 
the wind tends to flow into star's magnetic equatorial plane, producing shocks and
prominence-like structures; farther away the flow can carry close loops of magnetic fields,
much like in the Solar wind.
Magnetic spots on the surface of hot stars might also be responsible for 
variable magnetic stellar
wind activity and clumping \citep{2011A&A...534A.140C}. 
In a detached binary system, the wind structure can be even more complicated, 
as the OB-companion is likely to be non corotating with the orbital period and may have 
the spin axis inclined to the orbital plane (due to, e.g., the natal kick 
acquired by the NS during the formation in supernova explosion).   
Therefore, it is possible that the wind material lost by the supergiant companions of SFXTs is non-magnetized for most of the time, and only sporadically it transports with it part of the star magnetic field.
The shell instability described above can be triggered by large-scale
magnetic field sporadically carried by the stellar wind of the optical OB companion.

It is also well known from Solar wind studies (see e.g. reviews \cite{2004PhyU...47R...1Z, lrsp-2013-2} and references therein) that the Solar wind patches carrying tangent magnetic fields 
has a  lower velocity (about $350$~km~s$^{-1}$) than the wind with radial magnetic fields 
(up to $\sim 700$~km s$^{-1}$). Fluctuations of the stellar wind density and velocity 
from massive stars are also known from spectroscopic observations \citep{2008A&ARv..16..209P}, 
with the typical velocity fluctuations up to $0.1\ v_\infty\sim 200-300$~km s$^{-1}$.  

The effect of the magnetic field carried by the stellar wind is twofold: first, 
it can trigger
rapid mass entry rate to the magnetosphere via magnetic reconnection in the magnetopause (the phenomenon well known in the dayside Earth magnetosphere, \cite{1961PhRvL...6...47D}), and second, 
the magnetized clumps with tangent magnetic field
have smaller velocity than unmagnetized 
ones (or carrying the radial field). The first factor, under certain conditions discussed below, can increase the plasma fall velocity in the shell from inefficient, radiative-cooling controlled settling accretion 
with $f(u)_{rad}\sim 0.03-0.1$, 
up to the maximum possible free-fall velocity with $f(u)=1$.
In other words, during the bright flare the subsonic 
settling accretion transforms into the supersonic Bondi accretion.
The second factor (slower wind velocity in magnetized clumps with tangent 
magnetic field) 
strongly increases the Bondi radius $R_B\propto v_w^{-2}$
and the corresponding Bondi mass accretion rate $\dot M_B\propto v_w^{-3}$. 
Thus, we suggest that 
should a magnetized density clump be captured, the magnetic reconnection 
can trigger the shell instability around the NS magnetosphere.

To substantiate these factors, let us write down the mass accretion rate onto the NS in the unflaring
(low-luminosity) state as $\dot M_{a,low}=f(u) \dot M_B$
with $f(u)$ given by formula (\ref{furad}) 
and $\dot M_B\simeq \pi R_B^2 \rho_w v_w $.
Eliminating the wind density $\rho_w$ using the mass continuity equation written for the 
spherically symmetric stellar wind from the optical star with power $\dot M_o$ and by assuming  
circular binary orbit, we arrive at 
$
\dot M_B\simeq \frac{1}{4}\dot M_o \myfrac{R_B}{a}^2\,.
$
Next, let us exploit 
the well-known relation for the radiative wind mass-loss rate from massive hot stars
$%\beq{Lamers}
\dot M_o\simeq \epsilon \frac{L}{cv_\infty}
$%\eeq
where $L$ is the optical star luminosity, $v_\infty$ is the stellar wind velocity at infinity,
typically 2000-3000 km s$^{-1}$ for OB stars, $\epsilon\simeq 0.4-1$ is the efficiency factor \citep{1976A&A....49..327L} (in the numerical estimates below we shall assume $\epsilon=0.5$). 
It is also possible to reduce the luminosity $L$ of a
massive star  to its mass $M$ using 
the phenomenological relation $(L/L_\odot)\approx 19 (M/M_\odot)^{2.76}$ (see e.g. \cite{2007AstL...33..251V}). Combining above equations and using
third Kepler's law to express the orbital separation $a$ through the binary period $P_b$, we find for the X-ray luminosity of SFXT in the non-flaring state 
\begin{eqnarray}
\label{Lxlow}
L_{x,low}\simeq & 5\times 10^{35} [\hbox{erg~s}^{-1}] f(u) 
\myfrac{M}{10 M_\odot}^{2.76-2/3} \nonumber\\
&\myfrac{v_\infty}{1000 \mathrm{km~s}^{-1}}^{-1}
\myfrac{v_w}{500 \mathrm{km~s}^{-1}}^{-4}\myfrac{P_b}{10 \mathrm{d}}^{-4/3}\,,
\end{eqnarray}
which for $f(u)\sim 0.03-0.1$ 
corresponds to the typical low-state luminosities of SFXTs $\sim 10^{34}$~erg~s$^{-1}$. 

It is straightforward to see that the transition from the low state (subsonic accretion with 
slow magnetospheric entry rate $f(u)\sim 0.03-0.1$) to the supersonic free-fall Bondi accretion 
with $f(u)=1$ due to the magnetized stellar wind with the velocity decreased by  factor two, for example, would lead to a flare luminosity of $L_{x,flare}\sim (10\div 30)\times 2^5 L_{x,low}$. This shows that 
the dynamical range of SFXT bright flares ($\sim 300-1000$) can be naturally reproduced by the proposed mechanism.

For magnetic field reconnection to occur, the time the magnetized plasma spends 
near the magnetopause should be at least comparable to 
the reconnection time, $t_r\sim R_A/v_r$, where
$v_r$ is the magnetic reconnection rate, which is difficult to assess from the first principles
\citep{2009ARA&A..47..291Z}.
For example, in the Petschek fast reconnection model $v_r=v_A(\pi/8\ln S)$, where $v_A$ is the 
Alfv\'en speed and $S$ is the Lundquist number (the ratio of the Ohmic diffusion time to 
the Alfv\'en time); for typical conditions near NS magnetospheres we
find $S\sim 10^{28}$ and $v_r\sim 0.006 v_A$. In real astrophysical plasmas 
the large-scale magnetic reconnection rate can be a few times as high, 
$v_r\sim 0.03-0.07 v_A$ \citep{2009ARA&A..47..291Z}, and, guided by phenomenology, we can parametrize it as $v_r=\epsilon_r v_A$ with $\epsilon_r\sim 0.01-0.1$. The longest time-scale the plasma penetrating into the magnetosphere spends near the magnetopause
is the instability time, $t_{inst}\sim t_{ff}(R_A)f(u)_{rad}$ \citep{2012MNRAS.420..216S}, so the 
reconnection can occur if 
$t_r/t_{inst}\sim (u_{ff}/v_A)(f(u)_{rad}/\epsilon_r)\lesssim 1$. As near 
$R_A$ (from its definition) $v_A\sim u_{ff}$, we arrive at $f(u)_{rad}\lesssim\epsilon_r$ as
the necessary reconnection condition. By Eq. (\ref{furad}), it is satisfied only 
at sufficiently low X-ray luminosities, pertinent to 'quiet' SFXT states. 
\textit{This explains why in HMXBs with convective shells and higher luminosity
(but still smaller than $4\times 10^{36}$~erg~s$^{-1}$, where the settling accretion is possible), 
%%
%
%We note that due to a very complex nature of magnetic reconnection, 
%even when the magnetized plasma
%is accreted, 
the reconnection from magnetized plasma accretion will not
lead to the shell instability, but only 
to temporal establishing of the 'strong coupling regime' 
of angular momentum transfer through the shell, as 
discussed in \citet{2012MNRAS.420..216S}.} 
Episodic strong spin-ups, not associated with strong 
X-ray flux variations, as observed in GX 301-2, 
can be manifestations of such 'failed' 
reconnection-induced shell instability
\footnote{See \cite{2012ApJ...753....1I} for an
alternative explanation of strong spin-ups in GX 301-2 in the frame of another theory 
of magnetic accretion.}.

Therefore, it seems likely that the key (distinctive) difference between 
steady HMXBs like Vela X-1, GX 301-2 (showing only moderate flaring activity) and SFXTs is
that in the first case the effects of possible magnetized stellar winds from optical OB-companions
are insignificant (basically due to rather high accretion rate),
while in SFXTs with lower 'steady' X-ray luminosity, 
large-scale magnetic field, sporadically carried by the wind clumps, 
can trigger SFXT flaring activity via magnetic reconnection near the magnetospheric boundary. 
The observed power-law SFXT flare distributions, discussed in \citet{PaizisSidoli2014},
with respect to the log-normal distributions for classical HMXBs \citep{2010A&A...519A..37F}, 
may be related to the properties of magnetized stellar wind and physics of its interaction 
with the NS magnetosphere.

%quiet subsonic settling accretion.

\section{Discussion and Conclusions}

%%%%%%%%%%%%%%%%%%%%%%%%%%%%%%%%%%%%%%%%%%%%%%%%%%%%%%%%%%%%%%%%%%%%% Fig 3
\begin{figure}
\begin{center}
\includegraphics[width=0.45\textwidth]{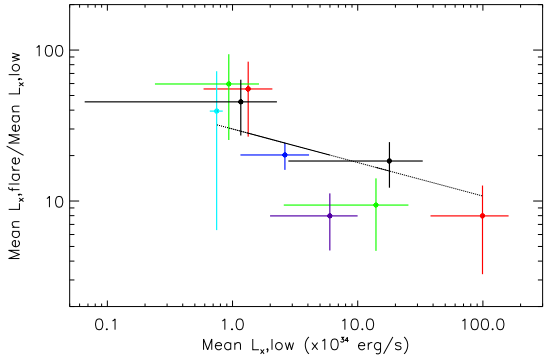}
%\parbox[t]{0.45\textwidth}
{\caption{The mean dynamic range of SFXT bright flares $L_{x,flare}/L_{x,low}$ as a function of low (unflaring) X-ray luminosity for sources from Fig. \ref{fig:energy} 
with errors shown as explained in text. Sources are colored as in Fig.~1. The \textit{x-axis} is in units of $10^{34}$~erg~s$^{-1}$. 
The straight line shows 
the expected dependence (\ref{dramp}).
}
\label{fig:dynrange}
}
\end{center}
\end{figure}
%%%%%%%%%%%%%%%%%%%%%%%%%%%%%%%%%%%%%%%%%%%%%%%%%%%%%%%%%%%%%%%%%%%%%%%%

The actually observed dynamic range of bright flares in SFXTs may be higher than 1000. 
The ratio between $L_x$ at the peak of some bright flares, which can be as high 
as a few $\times 10^{36}$~erg~s$^{-1}$, sometimes up to $10^{37}$~erg~s$^{-1}$, 
and the $L_x$ in the non-flaring state (the weakest $L_x$ ever measured in a couple of SFXTs is $L_{x,low}=10^{32}$~erg s$^{-1}$), can reach $10^5$. This very high dynamic range is observed, for example, in IGR 17544-2619 (\citealt{zand2005},  \citealt{Rampy2009}). 

Flares in SFXTs exhibit a rich phenomenology, including single outbursts (like in IGR  J18410-0535)
or several successive flares on top of a varying base flux (like in IGR J11215--5952, 
\cite{2009ApJ...696.2068R}). In the quasi-spherical accretion model discussed above, 
the dependence of the quiescent luminosity  
on the wind velocity is very high, $L_{x,low}\sim v_w^{-5}$ (see Eq. (\ref{Lxlow}) above). 
This means that in a steady state  
smooth velocity variations by a factor of two (for example, due to
high orbital eccentricity)
could change the X-ray luminosity between the flares 
by a factor of 30. If the magnetized clumps are sporadically appear in the stellar wind, flares on top of smoothly varying $L_{x,low}$ are expected to 
occur with the amplitude 
\begin{equation}
\label{dramp}
L_{x,flare}/L_{x,low}\sim f(u)_{rad}^{-1}\sim 30L_{34}^{-2/9}
\end{equation}
This may be the case of IGR J11215--5952 \citep{2009ApJ...696.2068R}. Thus, the stronger wind velocity variations in low-luminosity sources could produce higher dynamic range of flares.

In Fig. \ref{fig:dynrange} we plot the mean dynamic range of bright SFXT flares $L_{x,flare}/L_{x,low}$ as a function of $L_{x,low}$ for sources from Fig. \ref{fig:energy}.  
The errors on the \emph{x-axis} (mean $L_{x,low}$)  
are calculated as in Fig.~2, while the uncertainties on 
\emph{y-axis},  the ratio of the luminosities, 
are obtained considering the standard deviation 
from the mean flare luminosity (for $L_{x,flare}$) and the statistical errors for 
$L_{x,low}$. 
%The formal rms fit to these points gives the dependence $L_{x,flare}/L_{x,low}=
%(34\pm9)L_{34}^{-0.53\pm  0.18}$, in qulaitative ($\sim 2\sigma$) agreement 
The solid line shows the expected model dependence according to  Eq. (\ref{dramp}).

The proposed mechanism of SFXT bright flares is of course highly simplified. For example, the binary
orbit can be eccentric, the properties of magnetized winds from OB-supergiants can be anisotropic (see, e.g., model calculations in \cite{2002ApJ...576..413U} and other works of those authors), 
the shells can be not spherically symmetric, the photon field is definitely not spherically symmetric, etc. Nevertheless, the naturalness of the energy and time scales of bright flares, their dynamic range and the dependence of these properties 
on the unflaring X-ray luminosity seem to be highly suggestive. The observed spread in SFXT flare luminosity and energy distribution (\cite{PaizisSidoli2014} and Fig. \ref{fig:energy} in the present paper) can be due to the complexity and diversity of the magnetic field reconnection near 
the magnetopause and variations of the stellar wind properties. 
%A distinctive
%feature of the model could be the possible flux of non-thermal particles generated 
%in the magnetic field reconnection. However, un upper limit of their power, by assuming
%$\sim 10\%$ non-thermal particle generation in magnetic reconnection, is much less than 
%$\mu^2/R_A^3\sim 10^{33}$~erg s$^{-1}$, and is currently unobservable. 

We conclude that the SFXT bright flares can be caused by sporadic transitions between 
different regimes of accretion in a quasi-spherical
shell around a slowly rotating magnetized neutron star. The non-flaring steady 
states of SFXTs with low X-ray luminosity $L_{x,low}\sim 10^{32}-10^{34}$~erg~s$^{-1}$ may be  
associated with settling subsonic accretion mediated by 
ineffective radiative plasma cooling near the magnetospheric boundary. 
This regime is supported by observations of the 
X-ray pulse profile phase shift below some X-ray luminosity in 
XMM-Newton observations of the SFXT IGR~J11215--5952 \citep{Sidoli2007}. 
At this stage, the accretion rate onto the neutron star is suppressed 
by a factor of $\sim 30$ relative to the Bondi-Hoyle-Littleton value. 
Changes in the local wind velocity and density due to, e.g., clumps, can only slightly increase the mass accretion rate
(a factor of $\sim 10$) bringing the system into the Compton cooling dominated regime and led to the production of moderately bright flares ($L_x\lesssim 10^{36}$~erg/s).
To interpret the brightest flares ($L_x>10^{36}$~erg/s) displayed by the SFXTs within the quasi-spherical settling accretion regimes,
we propose that a larger increase in the mass accretion rate can be produced by 
sporadic capture of magnetized stellar wind plasma. 
Such episodes should not be associated with specific binary orbital phases, as 
observed in e.g. IGR J17544-2619 \citep{2014MNRAS.439.2175D}. 
At sufficently low accretion rates, 
magnetic reconnection 
can enhance the magnetospheric plasma entry rate, resulting in copious production of X-ray photons,
strong Compton cooling and ultimately in unstable 
accretion of the entire shell. 
A bright flare develops on the free-fall time
scale in the shell, and the typical energy released in an SFXT bright flare corresponds to 
the entire mass of the shell. This view is consistent with 
the energy released in SFXT bright flares ($\sim 10^{38}-10^{40}$~ergs),
their typical dynamic range ($\sim 100$), and with the observed dependence of these characteristics on the average unflaring X-ray luminosity of SFXTs. 
Thus the flaring behaviour of SFXTs, as opposed to steady HMXBs, may be primarily related to 
their low X-ray luminosity in the settling accretion regime, 
allowing sporadic magnetic reconnection to occur during 
magnetized plasma entry the NS magnetosphere.

\section{Acknowledgements}
The authors thank the anonymous referee for notes and A. Lutovinov and
B.V. Somov for discussions.
NSh acknowledges the Russian Science Foundation grant 14-02-00146. 
KP was supported by the RFBR grant 14-02-00657a. 
AP and LS acknowledge the Italian Space Agency financial  support {\it INTEGRAL}  
ASI/INAF agreement n. 2013-025.R.0.

\bibliographystyle{mn2e}
\expandafter\ifx\csname natexlab\endcsname\relax\def\natexlab#1{#1}\fi
\bibliography{wind}

\end{document}